\RequirePackage{ifpdf}
\ifpdf 
\documentclass[pdftex]{sigma}
\else
\documentclass{sigma}
\fi

\usepackage{bm} 

    \DeclareMathOperator\GF{GF}

    \DeclareMathOperator\GQ{GQ}

\newcommand{\cW}{{\mathcal W}} 
\newcommand{\cQ}{{\mathcal Q}}

\newcommand{\bP}{{\mathbb P}}
\newcommand{\bR}{{\mathbb R}}
\newcommand{\bZ}{{\mathbb Z}}

\newcommand{\vG}{{\bm G}}
\newcommand{\vC}{{\bm C}}
\newcommand{\vK}{{\bm K}}
\newcommand{\vN}{{\bm N}}
\newcommand{\vM}{{\bm M}}
\newcommand{\vS}{{\bm S}}
\newcommand{\vT}{{\bm T}}

{\theoremstyle{definition}
\newtheorem{cond}{Condition}}

\newenvironment{abc}{%
\begin{enumerate}}{\end{enumerate}}

\begin{document}

\renewcommand{\PaperNumber}{096}

\FirstPageHeading

\ShortArticleName{Factor-Group-Generated Polar Spaces and (Multi-)Qudits}

\ArticleName{Factor-Group-Generated Polar Spaces\\ and (Multi-)Qudits}

\Author{Hans HAVLICEK~$^{\dag\S}$, Boris ODEHNAL~$^\dag$ and Metod SANIGA~$^{\ddag\S}$}

\AuthorNameForHeading{H.~Havlicek, B.~Odehnal and M.~Saniga}

\Address{$^\dag$~Institut f\"{u}r Diskrete Mathematik und Geometrie, Technische
Universit\"{a}t Wien,\\
\hphantom{$^\dag$}~Wiedner Hauptstra{\ss}e 8--10/104, A-1040 Wien, Austria}
\EmailD{\href{mailto:havlicek@geometrie.tuwien.ac.at}{havlicek@geometrie.tuwien.ac.at}, \href{mailto:boris@geometrie.tuwien.ac.at}{boris@geometrie.tuwien.ac.at}}
\URLaddressD{\url{http://www.geometrie.tuwien.ac.at/havlicek/}}
\URLaddressD{\url{http://www.geometrie.tuwien.ac.at/odehnal/}}

\Address{$^\ddag$~Astronomical Institute, Slovak Academy of Sciences,\\
\hphantom{$^\ddag$}~SK-05960 Tatransk\'{a} Lomnica, Slovak Republic}
\EmailD{\href{mailto:msaniga@astro.sk}{msaniga@astro.sk}}
\URLaddressD{\url{http://www.astro.sk/~msaniga/}}

\Address{$^{\S}$~Center for Interdisciplineary Research (ZiF),
University of Bielefeld,\\
\hphantom{$^{\S}$}~D-33615 Bielefeld, Germany}

\ArticleDates{Received August 19, 2009, in f\/inal form October 02, 2009;  Published online October 13, 2009}

\Abstract{Recently, a number of interesting relations have been discovered
between genera\-lised Pauli/Dirac groups and certain f\/inite geometries. Here, we
succeeded in f\/inding a~general unifying framework for all these relations. We
introduce gradually necessary and suf\/f\/icient conditions to be met in order to
carry out the following programme: Given a group~$\vG$, we f\/irst construct
vector spaces over~$\GF(p)$, $p$ a prime, by factorising $\vG$ over appropriate
normal subgroups. Then, by expressing~$\GF(p)$ in terms of the commutator
subgroup of~$\vG$, we construct alternating bilinear forms, which ref\/lect
whether or not two elements of~$\vG$ commute. Restricting to $p=2$, we search
for ``ref\/inements'' in terms of quadratic forms, which capture the fact whether
or not the order of an element of $\vG$ is $\leq 2$. Such
factor-group-generated vector spaces admit a natural reinterpretation in the
language of symplectic and orthogonal polar spaces, where each point becomes a
``condensation'' of several distinct elements of $\vG$. Finally, several
well-known physical examples (single- and two-qubit Pauli groups, both the real
and complex case) are worked out in detail to illustrate the f\/ine traits of the
formalism.}


\Classification{20C35; 51A50; 81R05}

\section{Introduction}

The purpose of this paper is to establish the most general formal setting for
reformulating, whenever possible, basic properties of groups in terms of vector
spaces, alternating bilinear forms, quadratic forms and associated projective
and polar spaces. As far as we know, the f\/irst outline of such an analysis can
be tracked back in the textbook of Huppert \cite{huppert-67}, when addressing
the so-called {\em extra-special groups}; however, the assumptions made there
were rather specif\/ic and no f\/inite geometry was explicitly mentioned. Another
treatment of the issue, with important physical applications, was given by Shaw
and his collaborators \cite{shaw-89,shaw-91, shaw-92,shaw-93, shaw-95,shaw+j-90}. These papers deal with the
{\it Dirac groups} and their relationship to projective spaces over $\GF(2)$.
They include also a detailed dictionary from group theory to f\/inite geometry
and \emph{vice versa} (see also \cite{gordon+n+j+m+s-94}). Being unaware of
these developments, Planat and Saniga and others set up a similar programme
\cite{planat+s+k-06a, saniga+planat-06a,saniga+planat-07b,
saniga+p+m-07a, saniga+planat-07a, havlicek+saniga-07a,
saniga+p+p+h-07a, planat+baboin-07a,planat+b+s-08a,
havlicek+saniga-08a,planat+saniga-08a, saniga+p+p-08a,
levay+s+v-08a,levay+s+v+p-08a} (see also~\cite{rau-09a}), and
discovered various kinds of f\/inite geometry behind the generalised Pauli groups
of specif\/ic f\/inite-level quantum systems, their results being put into a more
general context by Koen Thas~\cite{thas-09z} ($p=2$) and~\cite{thas-09y}
($p>2$); these works, however, focussed uniquely on symplectic case
(alternating bilinear forms), leaving the importance of quadratic forms simply
unnoticed. In what follows we shall not only f\/ill this gap, but develop the
theory to such an extent that the links between the above-mentioned approaches
become clearly visible and, at the same time, dif\/ferences between some closely
related f\/inite groups (e.g., between the real and complex two-qubit Pauli
groups) will be revealed and properly understood.

\section{Preliminaries}

We f\/irst collect some notions which will be used throughout the paper:

Let $(\vG,\cdot)$ be a group with neutral element $e$. Given a set
$\vM\subseteq \vG$ we denote by $\langle \vM\rangle$ the subgroup of $\vG$
generated by $\vM$. Also, we let
\begin{equation*}
    \vM^{(m)}:=\{x^m \mid x\in \vM\}\quad \mbox{for all} \ \ m\in\bZ.
\end{equation*}
The commutator of $a,b\in \vG$ is written as $ [a,b]:=aba^{-1}b^{-1}$. The
commutator group (derived group) $[\vG,\vG]=:\vG'$ is the subgroup of $\vG$
which is generated by all commutators. The centre of $\vG$ is written as
$Z(\vG)$.

Furthermore, let $p$ be a f\/ixed prime. We denote the Galois f\/ield with $p$
elements by $\GF(p)=\bZ/(\bZ p)$. We shall always use $0,1,\ldots,p-1\in\bZ$ as
representatives for the elements of $\GF(p)$. Vector spaces over $\GF(p)$ have
a series of rather simple, but nevertheless noteworthy properties which are not
shared by vector spaces over arbitrary f\/ields. If $(V,+)$ is vector space over
$\GF(p)$ then
\begin{equation}\label{eq:mv}
    mv = \underbrace{v+v+\cdots+v}_{m} \quad \mbox{for all} \ \ m\in\GF(p),\ v\in V.
\end{equation}
So the additive group $(V,+)$ or, more precisely, $V$ as a $\bZ$-module,
determines the structure as a vector space over $\GF(p)$ in a \emph{unique}
way. In particular, we have
\begin{equation}\label{eq:pv=0}
    \underbrace{v+v+\cdots+v}_{p}=o \quad \mbox{for all} \ \ v\in V,
\end{equation}
where $o$ denotes the zero element of $V$. Consequently, any subgroup of $V$ is
also a (vector) subspace. Furthermore, any additive mapping of vector spaces
over $\GF(p)$ is also linear; see, among others, \cite{kirsch-78} and
\cite{mayr-79}. Conversely, a commutative group $(V,+)$ satisfying
(\ref{eq:pv=0}) can be turned into a vector space over $\GF(p)$ by def\/ining the
product of $m\in\GF(p)$ and $v\in V$ by (\ref{eq:mv}).

\section[Vector spaces over $\GF(p)$]{Vector spaces over $\boldsymbol{\GF(p)}$}

We aim at constructing vector spaces over $\GF(p)$ by factorising $\vG$ modulo
appropriate normal subgroups.
\par
Let $\vN\unlhd \vG$, i.e., $\vN$ is a normal subgroup of $\vG$. The factor
group $\vG/\vN$ is commutative if, and only if, the commutator group satisf\/ies
$\vG'\leq \vN$. Furthermore, $\vG/\vN$ is isomorphic to the additive group of a
vector space over $\GF(p)$ if, and only if, it satisf\/ies the following
condition:

\begin{cond}\label{c:el-abel}
$\vN$ is a normal subgroup of $\vG$ which contains the commutator subgroup
$\vG'$ and the set $\vG^{(p)}$ of $p$th powers.
\end{cond}

\begin{remark}\label{rem:normal}
Let $\vN\leq \vG$ be a \emph{subgroup\/} of $\vG$ satisfying $\vG'\leq \vN$. We
recall that $\vN$ is a \emph{normal subgroup\/} of $\vG$ in this case, since
for all $a\in \vN$ and all $x\in \vG$ we have $xax^{-1}=[x,a]a\in \vN$. This
means that Condition~\ref{c:el-abel} can be relaxed by omitting the word
``normal''.
\end{remark}

\begin{remark}\label{rem:komplexprod}
The complex product $\vG'\vG^{(p)}=\{xy \mid x\in\vG',\, y\in \vG^{(p)}\}$ is
easily seen to be a~subgroup of $\vG$. Thus, by Remark~\ref{rem:normal}, we
have
\begin{equation}\label{eq:komplexprod}
    \vG'\vG^{(p)}=\langle \vG'\cup \vG^{(p)}\rangle \unlhd \vG.
\end{equation}
\end{remark}
\begin{remark}
The case $p=2$ deserves particular mention. Here Condition \ref{c:el-abel} can
be further relaxed by deleting the condition $\vG'\leq \vN$, because
$\vG^{(2)}\subseteq \vN$ implies that all elements of $\vG/\vN$ have order one
or two, which in turn guarantees the commutativity of $\vG/\vN$.\footnote{A
group of prime exponent $p>2$ need not be commutative. For example, the set of
upper triangular $3\times 3$ matrices over $\GF(p)$ with $1s$ along the
diagonal is a non-commutative group of exponent $p$ under matrix multiplication
for $p>2$.}
\end{remark}

\par
We assume until further notice that Condition \ref{c:el-abel} holds. Then we
let
\begin{equation*}
    (V,+):=(\vG/\vN,\cdot),
\end{equation*}
i.e., the composition in $V$ will be written additively, and we consider $V$
as a vector space over $\GF(p)$ in accordance with (\ref{eq:mv}).

It is an easy exercise to express notions from the vector space $V$ (like
linear dependence, dimension, etc.) in terms of the factor group $\vG/\vN$. For
example, a linear combination $\sum_{i=1}^{k} m_i v_i$ with $m_i\in\GF(p)$,
$v_i=x_i\vN$ and $x_i\in \vG$ translates into $x_1^{m_1}x_2^{m_2}\cdots
x_k^{m_k}\vN$. The factors in this product may be rearranged in any order. The
set of all subspaces of $V$ is precisely the set
\begin{equation}\label{eq:unterraeume}
    \{ \vS/\vN \mid \vN\leq\vS\leq \vG \}.
\end{equation}
The factor spaces of $V$ have the form $V/(\vS/\vN)$, with $\vS$ as above.
There exists the canonical isomorphism (of vector spaces)
\begin{equation*}
    \vG/\vS \to (\vG/\vN)/(\vS/\vN) : \ x\vS \mapsto (x\vN)(\vS/\vN)
\end{equation*}
by the homomorphism theorem. Therefore, up to the canonical identif\/ication
\begin{equation}\label{eq:ident}
    \vG/\vS \equiv (\vG/\vN)/(\vS/\vN)=V/(\vS/\vN),
\end{equation}
the set of all factor spaces of $V$ is precisely the set
\begin{equation*}
    \{ \vG/\vS \mid \vN\leq \vS \leq \vG \}.
\end{equation*}
The identif\/ication (\ref{eq:ident}) will frequently be used in what follows. If
$V$ is f\/inite then $\# V = p^d$, where $d$ is the dimension of $V$. Hence in
this case the dimension of $V$ can be found by a simple counting argument.

We close this section with a complete description of all vector spaces arising
from our previous construction.
\begin{theorem}\label{thm:alle1}
Let $\vG$ be any group. Then the following assertions hold:
\begin{abc}\itemsep=0pt
\item\label{thm:alle1.a}
The subgroup $ \vN_0:=\vG'\vG^{(p)}$ is normal in $\vG$ and meets the
requirements of Condition~\emph{\ref{c:el-abel}}. Hence it yields the vector
space $V_0:=\vG/\vN_0$ over $\GF(p)$.

\item\label{thm:alle1.b}
The set of vector spaces $\vG/\vN$, where $\vN\unlhd\vG$ is subject to
Condition~\emph{\ref{c:el-abel}}, is precisely the set of all factor spaces of
$V_0$, up to the canonical identification $\vG/\vN \equiv V_0/(\vN/\vN_0) $
from \emph{(\ref{eq:ident})}.
\end{abc}
\end{theorem}

\begin{proof}
Ad (\ref{thm:alle1.a}): This is clear by Remarks~\ref{rem:normal}
and~\ref{rem:komplexprod}.

Ad (\ref{thm:alle1.b}): A subgroup $\vN \leq \vG$ satisf\/ies
Condition~\ref{c:el-abel} if, and only if, $\vN_0\leq\vN$. Under these
circumstances the canonical identif\/ication from (\ref{eq:ident}) can be applied
to $\vG/\vN$. This establishes the result.
\end{proof}
The previous result can be rephrased as follows: Our construction yields (to
within isomorphism) precisely the homomorphic images of the vector space $V_0$.

Of course, in Theorem~\ref{thm:alle1} the trivial case $\vN_0=\vG$ may occur so
that $V_0$ turns out to be the zero vector space over $\GF(p)$. Take, for
example, $\vG$ as a cyclic group of prime order $\neq p$. At the other extreme,
if $\vG$ is a commutative group of index $p$ then $\vN_0=\{e\}$.

\section{The underlying f\/ield}\label{se:GF}

For our construction of an alternating bilinear form in Section~\ref{se:alt},
we shall need an interpretation of the Galois f\/ield $\GF(p)$ \emph{within the
group $\vG$ in terms of the commutator group\/} $\vG'$. The (multiplicative)
group $\vG'$ is isomorphic to the additive group of the Galois f\/ield $\GF(p)$
precisely when the following is satisf\/ied:
\begin{cond}\label{c:kommutator}
The commutator group $\vG'$ has order $p$.
\end{cond}
This is due to the fact that any two groups of order $p$ are cyclic and hence
isomorphic. Condition~\ref{c:kommutator} is very restrictive, in sharp contrast
to Condition~\ref{c:el-abel}.

\begin{remark}\label{r:nichtkomm}
Condition~\ref{c:kommutator} implies that $\vG$ is a \emph{non-commutative
group}, since $\vG'$ has to have more than one element.
\end{remark}
Let us assume until the end of this section that Condition~\ref{c:kommutator}
holds. For each generator $g$ of~$\vG'$ (viz.\ each element $g\in
\vG'\setminus\{e\}$) the mapping
\begin{equation}\label{eq:psi_g}
    \psi_g: \ (\vG',\cdot) \to \big({\GF(p)},+\big): \ g^m \mapsto m
    \quad \mbox{with} \ \ m\in\{0,1,\ldots,p-1\}
\end{equation}
is an isomorphism of groups. Given a generator $\tilde g\in \vG'$ there exists
an element $k\in\{1,\ldots,{p-1}\}$ such that $g=\tilde g^k$, whence
\begin{equation*}
    (\psi_{\tilde g}\circ\psi_g^{-1})( m ) = k m \quad \mbox{for all} \ \ m\in\GF(p).
\end{equation*}
Therefore, loosely speaking, $\vG'$ could be identif\/ied with $\GF(p)$ \emph{up
to a non-zero scalar} \mbox{$k\!\in\!\GF(p)$}. In fact, Condition~\ref{c:kommutator} just
guarantees that $\vG'$ is a one-dimensional vector space over~$\GF(p)$, but it
does not provide a unique way to identify $\vG'$ with $\GF(p)$ unless $p=2$.
Examples of groups satisfying Condition~\ref{c:kommutator} will be exhibited in
Section~\ref{se:examp}.

\begin{remark} If Conditions~\ref{c:el-abel} and \ref{c:kommutator} are
satisf\/ied then, taking into account $\psi_g^{-1}(m)=g^m$ and $v=x\vN$ for some
$m\in\GF(p)$ and some $x\in \vG$, it would be \emph{incorrect\/} to calculate
the product~$m v$ in terms of the factor group $\vG/\vN$ as $(g^m\vN)
(x\vN)=g^m x\vN$. For example, $m=0$ and $v\neq o$ (zero vector) yield $0\cdot
v=o$, but $g^0x\vN=x\vN=v\neq o$. Observe that this applies even in the case
$p=2$, where there is just one possibility for choosing an isomorphism
$\psi_g$.
\end{remark}

\section{An alternating bilinear form}\label{se:alt}

Given a group $\vG$ and a normal subgroup $\vN\unlhd\vG$ satisfying
Condition~\ref{c:el-abel}, we want to turn the commutator mapping $
[\cdot,\cdot] : \vG\times \vG \to \vG'$ into a function which is well def\/ined
on $V\times V$. This amounts to requiring that for all $x,y\in \vG$ their
commutator $[x,y]$ does not change if $x$ is replaced by any element from the
coset $x\vN$ and likewise for $y$. For any $a\in \vN$ we have $[x,y]=[xa,y]$
if, and only if,
\begin{equation*}
     x y x^{-1} y^{-1} = x a y a^{-1} x^{-1} y^{-1}
\end{equation*}
or, equivalently, $a y = y a$. Since here $y\in \vG$ is arbitrary, this holds
precisely when $a\in Z(\vG)$. We are thus led to the following:

\begin{cond}\label{c:zentral}
The normal subgroup $\vN$ is contained in the centre of $\vG$.
\end{cond}
By virtue of this condition, we have indeed $[x,y]=[xa,yb]$ for all $x,y\in
\vG$ and all $a,b\in \vN$. However, there does not seem to be an obvious
meaning of the commutator group $\vG'$ for our vector space $V$. Hence we
assume until further notice that Conditions~\ref{c:el-abel},
\ref{c:kommutator}, and \ref{c:zentral} hold. Therefore
\begin{gather}\label{eq:123-kette}
    \vG'\vG^{(p)} \unlhd \vN\unlhd Z(\vG)\lhd\vG
\end{gather}
is satisf\/ied. Also, we choose an isomorphism $\psi_g$ according to
(\ref{eq:psi_g}). This allows us to def\/ine a~mapping
\begin{equation}\label{eq:[]_g}
     [\cdot,\cdot]_g : \ V\times V\to \GF(p) :  \ (v,w)=(x\vN,y\vN)\mapsto \psi_g ([x,y]),
\end{equation}
where $x,y\in \vG$. We collect now several basic properties of this mapping.

\begin{theorem}\label{thm:123}
Suppose that the group $\vG$ and the normal subgroup $\vN\unlhd\vG$ satisfy
Conditions~\emph{\ref{c:el-abel}},~\emph{\ref{c:kommutator}}, and
\emph{\ref{c:zentral}}. Also, let $g$ be a generator of the commutator group
$\vG'$. Then the following assertions hold:
\begin{abc}\itemsep=0pt
\item\label{thm:123.a}
The mapping $[\cdot,\cdot]_g$ given by \emph{(\ref{eq:[]_g})} is an alternating
bilinear form on the vector space $V=\vG/\vN$.

\item\label{thm:123.b}
Two elements $x,y\in \vG$ commute if, and only if, the corresponding vectors
$v=x\vN,w=y\vN\in V$ are orthogonal with respect to $[\cdot,\cdot]_g$, i.e.,
$[v,w]_g=0$.

\item\label{thm:123.c}
The bilinear form $[\cdot,\cdot]_g$ is non-zero and has the radical
$V^\perp=Z(\vG)/\vN$. Consequently, this form is non-degenerate if, and only
if, $\vN$ coincides with the centre of\/ $\vG$.

\end{abc}
\end{theorem}

\begin{proof}
Ad (\ref{thm:123.a}): Given $x,y\in \vG$ we let $v:=x\vN$ and $w:=y\vN$. Then
\begin{equation*}
    [v,v]_g=\psi_g ([x,x])=\psi_g (e)=0
\end{equation*}
and
\begin{equation}\label{eq:schiefsymm}
    [w,v]_g=\psi_g ([y,x])=\psi_g \big([x,y]^{-1}\big)=-[w,v]_g.
\end{equation}
Also, we obtain
\begin{gather}
    [v_1+v_2,w]_g
               =\psi_g \big( (x_1x_2)y(x_1x_2)^{-1}y^{-1}\big)
               =\psi_g \big( x_1x_2y x_2^{-1}x_1^{-1}y^{-1}\big)\nonumber\\
\phantom{[v_1+v_2,w]_g}{}  =\psi_g \big( x_1  \underbrace{ \big(x_2y x_2^{-1}y^{-1}\big) }_{\in\,\vG'} x_1^{-1} x_1 y  x_1^{-1}y^{-1}\big)
              =\psi_g \big(\big(x_2y x_2^{-1}y^{-1}\big)\big( x_1 y  x_1^{-1}y^{-1}\big)\big)\nonumber\\
\phantom{[v_1+v_2,w]_g}{}                =\psi_g \big([x_2,y]\cdot [x_1,y]\big)
               = [v_1,w]_g + [v_2,w]_g.\label{eq:bilinear}
\end{gather}
Here we used that $\vG'$ is f\/ixed elementwise under the inner automorphism
given by $x_1$ due to~(\ref{eq:123-kette}). The last equality follows, because
Condition~\ref{c:kommutator} forces $\vG'$ to be commutative. From
(\ref{eq:schiefsymm}) and (\ref{eq:bilinear}), the function $[\cdot,\cdot]_g$
is biadditive and therefore also bilinear. Hence the assertion follows.

Ad (\ref{thm:123.b}): This is immediate from the def\/inition of
$[\cdot,\cdot]_g$.

Ad (\ref{thm:123.c}): We noted already in Remark~\ref{r:nichtkomm} that $\vG$
is a non-commutative group. Consequently, the bilinear form $[\cdot,\cdot]_g$
is non-zero. Its radical is
\begin{equation*}
    V^\perp=\{v\in V\mid v\perp w \mbox{~for all~}w\in V\}.
\end{equation*}
We read of\/f from (\ref{thm:123.b}) that $V^\perp = Z(\vG)/\vN$ and the rest is
clear.
\end{proof}

Observe that the bilinear form $[\cdot,\cdot]_g$ has to be degenerate when
$\dim V$ is an odd integer. See Examples~\ref{exa:pauli16} and~\ref{exa:pauli64} in Section~\ref{se:examp}.

The previous result (\ref{thm:123.b}) about commuting elements does not depend
on the choice of the isomorphism $\psi_g$. Replacing $g$ by any generator
$\tilde g$ of the commutator group $\vG'$ changes the bilinear form
$[\cdot,\cdot]_g$ by a non-zero factor $k\in\GF(p)$, that is
$[\cdot,\cdot]_{\tilde g} = k[\cdot,\cdot]_g$. But the orthogonality relations
with respect to these two forms are identical. We could even rule out the
isomorphism~$\psi_g$ by considering the mapping $V\times V\to
\vG':(x\vN,y\vN)\mapsto [x,y]$. The proof of Theorem~\ref{thm:123} shows that
this is a non-zero alternating bilinear mapping of vector spaces over $\GF(p)$.
The interpretation of our results in terms of projective geometry will also
eliminate the explicit choice of an isomorphism $\psi_g$. See Section~\ref{se:polar}.

\par
We end with a complete description of all vector spaces and all alternating
bilinear forms arising from our construction from the above; cf.\
Theorem~\ref{thm:alle1}.

\begin{theorem}\label{thm:alle123}
Let $\vG$ be a group such that Condition~\emph{\ref{c:kommutator}} holds.
Furthermore, let at least one of the normal subgroups of $\vG$ satisfy
Conditions~\emph{\ref{c:el-abel}} and \emph{\ref{c:zentral}}. Choose $g\in
\vG'\setminus\{e\}$. Then the following assertions hold:
\begin{abc}\itemsep=0pt
\item\label{thm:alle123.a}
The subgroup $ \vN_0= \vG'\vG^{(p)}$ is normal in $\vG$ and meets the
requirements of Conditions~\emph{\ref{c:el-abel}} and \emph{\ref{c:zentral}}.
It yields the vector space $V_0=\vG/\vN_0$ over $\GF(p)$, the alternating
bilinear form $[\cdot,\cdot]_{g,0}$ on $V_0$, and the radical $V_0^\perp$.

\item\label{thm:alle123.b}
The set of vector spaces $\vG/\vN$, where $\vN$ is subject to
Conditions~\emph{\ref{c:el-abel}} and \emph{\ref{c:zentral}}, is precisely the
set of factor spaces $V_0/S$, where $S$ is any subspace of $V_0^\perp$, up to
the canonical identification from \emph{(\ref{eq:ident})}.

\item\label{thm:alle123.c}
In terms of the identification from \emph{(\ref{eq:ident})} the alternating
bilinear form $[\cdot,\cdot]_g$ on any vector space $\vG/\vN \equiv V_0/S$ as
in \emph{(\ref{thm:alle123.b})} is inherited from the bilinear form
$[\cdot,\cdot]_{g,0}$ on $V_0$.
\end{abc}
\end{theorem}

\begin{proof}
Ad (\ref{thm:alle123.a}): By the hypotheses of the theorem, $\vN_0\leq Z(\vG)$
holds, whence (\ref{thm:alle123.a}) is fulf\/illed.
\par
Ad (\ref{thm:alle123.b}): A subgroup $\vN\leq \vG$ satisf\/ies
Conditions~\ref{c:el-abel} and \ref{c:zentral} if, and only if,
$\vN_0\leq\vN\leq Z(\vG)$ which in turn is equivalent to
\begin{equation*}
    \vN_0\leq \vN \qquad \mbox{and}\qquad S = \vN/\vN_0 \leq Z(\vG)/\vN_0 = V_0^\perp.
\end{equation*}

Ad (\ref{thm:alle123.c}): The bilinear form $[\cdot,\cdot]_{g,0}$ induces a
well-def\/ined bilinear form on $V_0/S$ for any subspace $S\leq V_0^\perp$ via
$(v+S,w+S)\mapsto[v,w]_{g,0}$. This induced form coincides with
$[\cdot,\cdot]_g$ by its def\/inition.
\end{proof}

\section{A quadratic form}\label{se:quad}

We let $p:=2$ throughout this section. We exhibit a group $\vG$ and a normal
subgroup $\vN$ satisfying Conditions~\ref{c:el-abel} and \ref{c:kommutator},
but we do not yet assume Condition~\ref{c:zentral} to be fulf\/illed. So
$\vG'=\{e,g\}$, say, and $g=g^{-1}\neq e$. Hence the vector space $V=\vG/\vN$
and the (only) isomorphism $\psi_g: (\vG',\cdot)\to \big({\GF(2),+}\big)$ are
at our disposal. In the sequel the group
\begin{equation}\label{eq:K}
    \vK:=\{x\in Z(\vG)\mid x^2=e\}\leq Z(\vG)
\end{equation}
will play an important role.
\par
Our f\/irst aim is merely to def\/ine a mapping $\vG\to \GF(2)$ by the assignment
$x\mapsto \psi_g(x^2)$. This is possible if, and only if, the following holds:
\begin{cond}\label{c:x^2}
$\vG$ is a group such that all its squares belong to its commutator group,
i.e., $\vG^{(2)}\subseteq \vG'$.
\end{cond}
\begin{remark}\label{r:G2=G'}
We note that Conditions~\ref{c:kommutator} and \ref{c:x^2} imply
\begin{equation*}
    \vG^{(2)}= \vG',
\end{equation*}
since otherwise $\vG^{(2)}=\{e\}$ would force $\vG$ to be commutative, a
contradiction to Remark~\ref{r:nichtkomm}.
\end{remark}

We continue by demanding that also Condition~\ref{c:x^2} is satisf\/ied. Our
second aim is to f\/ind necessary and suf\/f\/icient conditions for the
mapping\footnote{We refrain from writing $Q_g$, since there is only one choice
for $g$, even though we maintain the notation $[\cdot,\cdot]_g$ from the
previous section.}
\begin{equation}\label{eq:Q}
    Q: \ V\to \GF(2): v=x\vN\mapsto \psi_g(x^2)
\end{equation}
to be well-def\/ined. This is the case if, and only if,
\begin{equation}\label{eq:Q_wohldef.1}
    x^2=xaxa \quad \mbox{for all} \ \ x\in \vG \quad \mbox{and all} \ \ a\in \vN.
\end{equation}
Let us consider f\/irst of all the particular case $x=e $ which yields the
necessary condition $a^2=e $ for all $a\in \vN$. As $\vN=\{e\}$ is impossible
due to $e\neq g\in \vN$, we continue by assuming the following to be true:

\begin{cond}\label{c:a-quadrat}
The normal subgroup $\vN\unlhd \vG$ has exponent $2$.
\end{cond}
Now, returning to the general case, we can use Condition~\ref{c:a-quadrat} to
rewrite (\ref{eq:Q_wohldef.1}) in the form
\begin{equation}\label{eq:Q_wohldef.2}
    x^2  = x^2(x^{-1}a^{-1}xa) \quad \mbox{for all} \ \ x\in \vG \quad \mbox{and all} \ \ a\in \vN,
\end{equation}
because $a=a^{-1}$. Cancelling $x^2$ shows that (\ref{eq:Q_wohldef.2}) holds
precisely when $\vN$ is in the centre of $\vG$. Hence, we also have to impose
Condition~\ref{c:zentral} to be valid.
\par
Conversely, with all f\/ive conditions at hand we obtain that the mapping $Q$ in
(\ref{eq:Q}) is indeed well def\/ined. We notice that under these circumstances
\begin{equation}\label{eq:1-5-kette}
    \{e\}=\vN^{(2)} \neq \{e,g\}=\vG'=\vG^{(2)}\unlhd \vN\unlhd\vK\unlhd Z(\vG)\lhd\vG
\end{equation}
is satisf\/ied. We are now in a position to describe the mapping $Q$ in detail.

\begin{theorem}\label{thm:1-5}
Suppose that the group $\vG$ and the normal subgroup $\vN\unlhd\vG$ satisfy
Conditions~\emph{\ref{c:el-abel}--\ref{c:a-quadrat}\/} for $p=2$. Then the
following assertions hold:
\begin{abc}\itemsep=0pt
\item\label{thm:1-5.a}
The mapping $Q: V \to \GF(2)$ given by \emph{(\ref{eq:Q})} is a quadratic form.
\item\label{thm:1-5.b}
The polar form of $Q$ equals to the alternating bilinear form given in
\emph{(\ref{eq:[]_g})}. Consequently, $Q$~is non-zero.
\item\label{thm:1-5.c}
The restriction of $Q$ to the radical $V^\perp$ is a linear form
$V^\perp\to\GF(2)$ with kernel \linebreak \mbox{$\vK/\vN\leq V^\perp$}. Hence either
$\vK/\vN=V^\perp$ or $\vK/\vN$ is a hyperplane of $V^\perp$.
\end{abc}
\end{theorem}

\begin{proof}
Ad (\ref{thm:1-5.a}) and (\ref{thm:1-5.b}): In order to show that $Q$ is a
quadratic form, we have to verify two conditions. Firstly, $Q(k v)=k^2 Q(v)$
for all $k\in\GF(2)$ and all $v\in V$. This follows from $Q(o)=\psi_g(e^2)=0$
for $k=0$ and is obviously true for $k=1$. Secondly, it remains to establish
that the mapping
\begin{equation*}
    V\times V\to \GF(2) : \ (v,w)\mapsto Q(v+w)-Q(v)-Q(w)
\end{equation*}
is biadditive and hence bilinear. Letting $v=x\vN$, $w=y\vN$ with $x,y\in \vG$
gives
\begin{equation}\label{eq:Q-polar}
    (xy)^2 x^{-2} y^{-2}= x^{-2} (xy)^2 y^{-2} = x^{-1}yxy^{-1}=[x^{-1},y].
\end{equation}
Here the f\/irst equation sign holds, because $\vG^{(2)}$ is a commutative group
by Remark~\ref{r:G2=G'}, which allows to rearrange squares. Application of
$\psi_g$ permits us to express (\ref{eq:Q-polar}) as
\begin{equation}\label{eq:Q-polar1}
    Q(v+w)-Q(v)-Q(w) = [-v,w]_g=[v,w]_g.
\end{equation}
Since $[\cdot,\cdot]_g$ is non-zero, so is $Q$. This completes the proof of
(\ref{thm:1-5.a}) and (\ref{thm:1-5.b}).

Ad~(\ref{thm:1-5.c}): The restriction of $Q$ to the radical
$V^\perp=Z(\vG)/\vN$ is additive by (\ref{eq:Q-polar1}). Hence $Q|V^\perp$ is a
linear form in $\GF(2)$. By its def\/inition, $Q|V^\perp$ vanishes precisely on
the set $\vK/\vN$, which is therefore all $V^\perp$, or one of its hyperplanes.
\end{proof}
\par
Our f\/inal result of this section is in the spirit of Theorems~\ref{thm:alle1}
and \ref{thm:alle123}:

\begin{theorem}\label{thm:alle1-5}
Let $\vG$ be a group such that Conditions~\emph{\ref{c:kommutator}} and
\emph{\ref{c:x^2}} hold for $p=2$. Furthermore, let at least one of the normal
subgroups of $\vG$ satisfy Conditions~\emph{\ref{c:el-abel}},
\emph{\ref{c:zentral}}, and \emph{\ref{c:a-quadrat}\/}. Then the following
assertions hold:
\begin{abc}\itemsep=0pt
\item\label{thm:alle1-5.a}
The normal subgroup $\vN_0=\vG'\vG^{(2)}=\vG'=\vG^{(2)}\unlhd\vG$ meets the
requirements of Conditions~\emph{\ref{c:el-abel}}, \emph{\ref{c:zentral}}, and
\emph{\ref{c:a-quadrat}\/}. It yields the vector space $V_0=\vG/\vN_0$ over
$\GF(p)$, the quadratic form~$Q_0$ on~$V_0$, and the subspace $\vK/\vN_0\leq
V_0^\perp$.

\item\label{thm:alle1-5.b}
The set of vector spaces $\vG/\vN$, where $\vN$ is subject to
Conditions~\emph{\ref{c:el-abel}}, \emph{\ref{c:zentral}},
and~\emph{\ref{c:a-quadrat}}, is precisely the set of factor spaces $V_0/S$,
where $S$ is any subspace of $\vK/\vN_0$, up to the canonical identification
from \emph{(\ref{eq:ident})}.

\item\label{thm:alle1-5.c}
In terms of the identification from \emph{(\ref{eq:ident})} the quadratic form
$Q$ on a vector space $\vG/\vN \equiv V_0/S$ as in \emph{(\ref{thm:alle1-5.b})}
is inherited from the quadratic form $Q_{0}$ on $V_0$.
\end{abc}
\end{theorem}

\begin{proof}
Ad (\ref{thm:alle1-5.a}): By the hypotheses of the theorem and
(\ref{eq:1-5-kette}), $\vG'=\vG^{(2)}=\vN_0\unlhd\vK \unlhd Z(\vG)$, whence
(\ref{thm:alle1-5.a}) is fulf\/illed.

\par
Ad (\ref{thm:alle1-5.b}): A subgroup $\vN\leq \vG$ satisf\/ies
Conditions~\ref{c:el-abel}, \ref{c:zentral} and \ref{c:a-quadrat} if, and only
if, $\vN_0\leq\vN\leq \vK $ which in turn is equivalent to
\begin{equation*}
    \vN_0\leq \vN \qquad \mbox{and}\qquad
    S = \vN/\vN_0 \leq \vK /\vN_0.
\end{equation*}

Ad (\ref{thm:alle1-5.c}): The quadratic form $Q_{0}$ induces a well-def\/ined
quadratic form on $V_0/S$ for any subspace $S\leq \vK /\vN_0$ via $v+S\mapsto
Q_0(v)$, because $Q_0(v+s)=Q_0(v)+Q_0(s)+[v,s]_{g_0}=Q_0(v)$ for all $s\in S$.
This induced form coincides with $Q$ by its def\/inition.
\end{proof}

Under the assumptions of Theorem~\ref{thm:alle1-5} suppose that $\vK< Z(\vG)$.
Then there exists a subgroup~$\vN$ with $\vK<\vN\leq Z(\vG)$, whence
Condition~\ref{c:a-quadrat} is violated, whereas
Conditions~\ref{c:el-abel}--\ref{c:zentral} are satisf\/ied. This means that the
vector space $\vG/\vN$ is endowed with an alternating bilinear form by
Theorem~\ref{thm:alle123}, but there exists no quadratic form on $\vG/\vN$ as
in Theorem~\ref{thm:alle1-5}; see Examples~\ref{exa:pauli16} and
\ref{exa:pauli64} in Section~\ref{se:examp}.

\section{Symplectic polar spaces}\label{se:polar}

Our results from the preceding sections allow a natural interpretation in terms
of projective geometry. Let $V$ be an $(n+1)$-dimensional\footnote{We restrict
ourselves to the f\/inite-dimensional case even though several results from below
could be carried over~-- \emph{mutatis mutandis}~-- to spaces of inf\/inite
dimension.} vector space over a f\/ield $F$. Recall that the \emph{points} of the
\emph{projective space} on $V$ are its one-dimensional subspaces (``rays
through the origin''). We write $\bP(V)$ for the set of all such points.
Likewise, each subspace $S$ of $V$ gives rise to a set $\bP(S)$ of points. If
$\dim S=k+1$ then $\bP(S)\subseteq\bP(V)$ is called a \emph{$k$-flat} or
\emph{$k$-dimensional projective subspace}. In particular, $\bP(V)$ is the only
$n$-f\/lat, i.e., its projective dimension is $n$. We use the familiar
terminology for low-dimensional f\/lats: \emph{lines}, \emph{planes}, and
\emph{solids\/} have projective dimension $1$, $2$, and $3$, respectively.
\emph{Hyperplanes\/} of $\bP(V)$ are those f\/lats $\bP(S)$ where $S$ has
codimension $1$ in $V$.

Assume now that $\big(V,[\cdot,\cdot]\big)$ is a \emph{symplectic vector
space}. So it is endowed with a non-degenerate alternating bilinear form
$[\cdot,\cdot]$, and $n+1=:2r$ is even. For each subset $W\subseteq V$ we
denote by~$W^\perp$ its \emph{orthogonal subspace}, i.e.\ the set of all
vectors in~$V$ which are orthogonal to every vector in~$W$. In particular,
$v^\perp$ is a subspace with codimension $1$ for each vector $v\in
V\setminus\{o\}$. In projective terms we obtain a \emph{null
polarity}\footnote{Other names for this mapping are \emph{symplectic polarity}
and \emph{null system}.}, i.e.\ the mapping which assigns to each point~$Fv$
its \emph{null hyperplane\/}~$\bP(v^\perp)$. More generally, one can associate
with each $k$-f\/lat~$\bP(S)$ the $(n-k-1)$-f\/lat~$\bP(S^\perp)$; it equals the
intersection of all hyperplanes $\bP(v^\perp)$, as $Fv$ ranges over all points
of~$\bP(S)$. A subspace $S\leq V$ is called \emph{totally isotropic} if $S\leq
S^\perp$. We use the same terminology for the f\/lat~$\bP(S)$. The
\emph{symplectic polar space} associated with $\big(V,[\cdot,\cdot])$ is the
point set $\bP(V)$ together with the set of all totally isotropic f\/lats. All
maximal totally isotropic f\/lats have projective dimension $r-1$. It is common
to denote this polar space by $\cW_{2r-1}(F)$ and, in particular
$\cW_{2r-1}(q)$ if $F=\GF(q)$ is a Galois f\/ield. For each $r$ and each $F$
there is a unique symplectic polar space to within isomorphisms; see
\cite{bue+c-95,cameron-00z}, and the references therein.

Two (not necessarily distinct) points $Fv$, $Fw$ of $\cW_{2r-1}(F)$ are said to
be \emph{conjugate\/} if $v\in w^\perp$ (or $w\in v^\perp$). In other words:
Two points are conjugate if one of them is in the null hyperplane of the other.
Two distinct points are conjugate precisely when they are on a common totally
isotropic line. Each point is self-conjugate.

It is now a straightforward task to establish a neat connection between our
previous results and symplectic polar spaces:

\begin{theorem}\label{thm:sympl}
Suppose that a group $\vG$ and its centre $Z(\vG)=:\vN$ satisfy
Conditions~\emph{\ref{c:el-abel}--\ref{c:zentral}} for some prime number $p$.
Furthermore, let $V:=\vG/Z(\vG)$ be finite and let an alternating bilinear form
$[\cdot,\cdot]_g$ be defined as in \emph{(\ref{eq:[]_g})}. Then the following
hold:
\begin{abc}\itemsep=0pt
  \item\label{thm:sympl.a}
$\big(V,[\cdot,\cdot]_g\big)$ gives rise to a finite symplectic polar space
$\cW_{2r-1}(p)$.
  \item\label{thm:sympl.b}
The totally isotropic flats of $\cW_{2r-1}(p)$ have the form
$\bP\big(\vC/Z(\vG)\big)$, where $\vC$ ranges over the set of all commutative
subgroups of $\vG$ which contain the centre $Z(\vG)$. In particular, the points
of $\cW_{2r-1}(p)$ have the form $\vC/Z(\vG)$, where $\vC:=\langle x\rangle
Z(\vG)$ and $x\in\vG\setminus Z(\vG)$.
  \item\label{thm:sympl.c}
Two elements $x,y\in\vG\setminus Z(\vG)$ commute if, and only if, the
corresponding points of $\cW_{2d-1}(p)$ are conjugate.
\end{abc}
\end{theorem}
\begin{proof}
Ad (\ref{thm:sympl.a}): By Theorem~\ref{thm:123}~(\ref{thm:123.c}), the form
$[\cdot,\cdot]_g$ is non-degenerate. Therefore $\dim V=:2r$ is even and the
assertion follows.
\par
Ad (\ref{thm:sympl.b}): By (\ref{eq:unterraeume}), any subspace of $V$ has the
form $\vS/Z(\vG)$ with $Z(\vG)\leq\vS\leq \vG$ and \emph{vice versa}. The
subspace $\vS/Z(\vG)$ is totally isotropic if, and only if, $[\cdot,\cdot]_g$
vanishes identically on~$\vS/Z(\vG)$. This holds precisely when the subgroup
$\vS$ is commutative. The points of $\cW_{2r-1}(p)$ are the one-dimensional
subspaces of $V$, i.e.\ the subgroups of $\vG/Z(\vG)$ which are generated by a
single element $xZ(\vG)$ with $x\in \vG\setminus Z(\vG)$. Hence they have the
asserted form.

Ad (\ref{thm:sympl.c}): This holds according to our def\/inition of
$[\cdot,\cdot]_g$ and the def\/inition of conjugate \linebreak points.
\end{proof}

The structure of the space $\cW_{2r-1}(p)$ from above ``is'' the structure of
commuting elements of~$\vG$. Note that any $x\in\vG\setminus Z(\vG)$ clearly
commutes with all powers of $x$ and with all elements of $Z(\vG)$. It is
therefore natural to ``condense'' the commutative subgroup $\langle x\rangle
Z(\vG)\leq \vG$ to a~single entity~-- a point of $\cW_{2r-1}(p)$. Also, it is
natural that all elements from the centre $Z(\vG)$ have no meaning for
$\cW_{2r-1}(p)$, as they commute with every element of $\vG$. We add in passing
that the polar space $\cW_{2r-1}(p)$ does not depend on the choice of the
generator $g$ of $\vG'$ which is used to def\/ine $[\cdot,\cdot]_g$.

\begin{remark}\label{rem:sympl.faktor}
The results from Theorem~\ref{thm:sympl} can be easily generalised to the
settings of Theorem~\ref{thm:123}. Under these circumstances the factor space
$V/V^\perp$ together with the alternating bilinear form, which is inherited
from $V$, takes over the role of the symplectic vector space from above. This
means that one gets a symplectic polar space in the projective space
$\bP(V/V^\perp)$. A $k$-f\/lat of $\bP(V/V^\perp)$ has, by def\/inition, the form
$\bP(S/V^\perp)$ with $V^\perp \leq S\leq V$ and $\dim (S/V^\perp) =k+1$. It
will be convenient to identify this f\/lat with the f\/lat $\bP(S)$ of the
projective space $\bP(V)$. From this point of view the f\/lats of
$\bP(V/V^\perp)$ are the f\/lats of $\bP(V)$ which contain $\bP(V^\perp)$. Note
that such a f\/lat now has \emph{two projective dimensions}. Its dimension with
respect to $\bP(V)$ is $\dim S-1$, while its dimension with respect to
$\bP(V/V^\perp)$ is $\dim (S/V^\perp) - 1$; see Example~\ref{exa:pauli16}.
\end{remark}

\section{Orthogonal polar spaces}\label{se:ortho}

In view of Section~\ref{se:quad} we adopt the following: Let $V$ be an
$(n+1)$-dimensional vector space over a f\/ield $F$ with characteristic $2$. Let
$Q:V\to F$ be a quadratic form and $[\cdot,\cdot]$ be its (alternating
bilinear) polar form. We assume $Q$ to be \emph{non-singular}, which means that
$Q(v)\neq 0$ for all non-zero vectors in the radical $V^\perp$. A subspace
$S\leq V$ is said to be \emph{singular\/} if $Q$ vanishes identically on $S$.
We use the same terminology for the f\/lat $\bP(S)$. The singular points of
$\bP(V)$ constitute a \emph{non-singular quadric\/} $\cQ$ of $\bP(V)$. The
\emph{orthogonal polar space} associated with $(V,Q)$ is the point set $\cQ$
together with all singular f\/lats \cite{bue+c-95, cameron-00z}. This
orthogonal polar space mirrors the ``intrinsic geometry'' of the quadric $\cQ$,
since the singular f\/lats are precisely those f\/lats which are entirely contained
in $\cQ$. For our purposes also the ``extrinsic geometry'', i.e.\ the points of
the ambient space $\bP(V)$ of\/f the quadric, will be important.

All maximal singular f\/lats of $\cQ$ have the same projective dimension $r-1$,
but the integer $r\geq 0$ depends heavily on the ground f\/ield $F$, the
dimension of $V$, and the quadratic form $Q$. We need here only the case
$F=\GF(2)$. It is well known that to within projective transformations only the
following cases occur \cite[p.~58]{bue+c-95},
\cite[pp.~121--126]{hirschfeld-98}:
\begin{center}\renewcommand\arraystretch{1.1}
\begin{tabular}{l|l|l|l|l}
 $n$    & $r-1$ & Symbol & $\#$ Point set & Name \\\hline\hline
 $2k$   & $k-1$ & $\cQ_{2k}(2)$   & $2^{2k} - 1$ & parabolic \\
 $2k+1$ & $k$   & $\cQ_{2k+1}^+(2)$ & $2^{2k+1} + 2^{k} - 1$ & hyperbolic \\
 $2k+1$ & $k-1$ & $\cQ_{2k+1}^-(2)$ & $2^{2k+1} - 2^{k} - 1$ & elliptic
\end{tabular}
\end{center}

For $n=2k$ the polar form of $Q$ is degenerate, $\dim V^\perp=1$. Hence
$V^\perp$ is a distinguished point, called \emph{nucleus}, in the ambient
projective space of $\cQ_{2k}(2)$, but it is not a point of $\cQ_{2k}(2)$.
Otherwise the polar form of $Q$ is non-degenerate. Below we use $\cQ(2)$ to
denote any of the quadrics from the above table.

\begin{theorem}\label{thm:ortho}
Suppose that a group $\vG$ and its subgroup $\vK=:\vN$ given by
\emph{(\ref{eq:K})} satisfy
Conditions~\emph{\ref{c:el-abel}--\ref{c:a-quadrat}} for $p=2$. Furthermore,
let $V:=\vG/\vK$ be finite and let a quadratic form $Q$ be defined as in
\emph{(\ref{eq:Q})}. Then the following hold:
\begin{abc}\itemsep=0pt
  \item\label{thm:ortho.a}
$Q$ gives rise to a non-singular quadric $\cQ(2)$ of~$\bP(V)$.
  \item\label{thm:ortho.b}
The totally singular flats of $\cQ(2)$ have the form $\bP\big(\vT/\vK)$, where
$\vT$ ranges over the set of all subgroups of $\vG$ which have exponent $2$ and
contain $\vK$. In particular, the points of $\cQ(2)$ have the form $\vT/\vK$,
where $\vT:=\langle x\rangle \vK$ with $x\in\vG\setminus \vK$ and $x^2=e$.
\end{abc}
\end{theorem}

\begin{proof}
Ad (\ref{thm:ortho.a}): By Theorem~\ref{thm:1-5}~(\ref{thm:1-5.c}), the
restriction of the quadratic form $Q$ to $V^\perp=Z(\vG)/\vK$ has the kernel
$\vK/\vK$. This is the zero-subspace of $V^\perp$, so that $Q$ is non-singular.

Ad (\ref{thm:ortho.b}): By (\ref{eq:unterraeume}), any subspace of $V$ has the
form $\vS/\vK$ with $\vK\leq\vS\leq \vG$ and \emph{vice versa}. The subspace
$\vS/\vK$ is singular if, and only if, $Q$ vanishes identically on $\vS/\vK$.
This holds precisely when the subgroup $\vS$ has exponent $2$. The points of
$\cQ(2)$ are the one-dimensional subspaces of~$V$, i.e.\ the subgroups of
$\vG/\vK$ which are generated by a single element $x\vK$ with $x\in
\vG\setminus \vK$ and $x^2=e$. Hence they have the asserted form.
\end{proof}

The structure of the polar space which is based on the quadric $\cQ(2)$ from
above ``is'' the structure of elements with order $2$ of the group $\vG$. Note
that for any $x\in\vG\setminus \vK $ with order $2$ the complex product
$\langle x\rangle \vK$ is a subgroup of $\vG$ with exponent $2$. It is
therefore natural to ``condense'' the subgroup $\langle x\rangle \vK\leq \vG$
to a single entity~-- a point of $\cQ(2)$. In our further discussion we have to
distinguish two cases:

If $n=2k+1$ is odd then the polar form of $Q$ is non-degenerate which implies
$\vK=Z(\vG)$. So the results of Theorems~\ref{thm:sympl} and~\ref{thm:ortho}
can be merged immediately. We obtain a symplectic polar space which is
``ref\/ined'' by an orthogonal one. The fact that subgroups of exponent $2$ are
commutative is mirrored in the fact that singular subspaces are totally
isotropic.

If $n=2k$ is even then $\vK\neq Z(\vG)$. The point $V^\perp=Z(\vG)/\vK$ is the
nucleus of the quad\-ric~$\cQ_{2k}(2)$. We have here the orthogonal polar space
given by $\cQ_{2k}(2)$ and the symplectic polar space $\cW_{2k-1}(2)$ which is
def\/ined in $\bP(V/V^\perp)$ according to Remark~\ref{rem:sympl.faktor}. It is
well known that these two spaces are isomorphic. An isomorphism is given by
``joining the quadric with its nucleus'': If~$\bP(S)$ is a singular subspace of
$\cQ_{2k}(2)$ then its join with the point $V^\perp$, i.e.\ $\bP(S+V^\perp)$,
is a~totally isotropic subspace of $\bP(V/V^\perp)$ and \emph{vice versa}. In
algebraic terms this gives the following bijection from the set of all
subgroups $\vT$ with exponent $2$ and $\vK\leq\vT\leq \vG$ onto the set of all
commutative subgroups $\vC$ with $Z(\vG)\leq\vC\leq\vG $:
\begin{equation*}
    \vT \mapsto \vC:=\vT Z(\vG).
\end{equation*}

\section{Illustrative examples from quantum theory}\label{se:examp}

\begin{example}\label{exa:pauli16}
We consider the \emph{complex Pauli matrices}
\begin{gather}\label{eq:pauli}
    \sigma_0:=\begin{pmatrix}1 & \hphantom{-}0\\0&\hphantom{-}1\end{pmatrix},\qquad
    \sigma_x:=\begin{pmatrix}0 & \hphantom{-}1\\1&\hphantom{-}0\end{pmatrix},\qquad
    \sigma_y:=\begin{pmatrix}0 & -i\\i&\hphantom{-}0\end{pmatrix},\qquad
    \sigma_z:=\begin{pmatrix}1 & \hphantom{-}0 \\0&-1\end{pmatrix}.
\end{gather}
The matrices $i^\alpha\sigma_\beta$ with $\alpha\in\{0,1,2,3\}$ and
$\beta=\{0,x,y,z\}$ constitute the \emph{Pauli group\/}  of order~$16$, which
is now our $\vG$. It acts on the two-dimensional complex Hilbert space of a
single qubit. In our terminology (with $p:=2$) we have $Z(\vG)=\{\pm \sigma_0,
\pm i\sigma_0\}$, $\vG'=\vG^{(2)}=\vK=\{\pm\sigma_0\}$ and $g=-\sigma_0$. The
group $\vG$ satisf\/ies Conditions~\ref{c:kommutator} and \ref{c:x^2}.

The normal subgroup $\vK=\vN_0$ satisf\/ies Conditions~\ref{c:el-abel},
\ref{c:zentral}, and \ref{c:a-quadrat}. The factor group $\vG/\vK$ has $2^3$
elements; it gives rise to a three-dimensional vector space $V_0$ over $\GF(2)$
as in Theorem~\ref{thm:123}~(\ref{thm:123.a}) with a degenerate alternating
bilinear form $[\cdot,\cdot]_{g,0}$. The projective space $\bP(V_0)$ is the
\emph{Fano plane}; see Fig.~\ref{abb:fano}.
The points of the Fano plane fall into three classes: The three
\textit{dark-shaded} points form a non-degenerate quadric $\cQ_2(2)$ (i.e.\ a
conic). They correspond to those elements of~$\vG\setminus\vK$ whose square is~$\sigma_0$ (i.e.\ Hermitian matrices). The three \textit{light-shaded} points
represent the elements of $\vG\setminus\vK$ whose square is~$-\sigma_0$ (i.e.\ skew-Hermitian matrices). The remaining point, which is depicted by a
\textit{double circle}, is the only point of~$\bP(V_0^\perp)$ or, in other
words, the nucleus of~$\cQ_2(2)$. It represents the matrices of~$Z(\vG)\setminus\vK$, which are also skew-Hermitian. The three lines through
the nucleus (bold-faced) are to be identif\/ied with the three ``points'' of the
symplectic polar space $\bP(V_0/V_0^\perp)\cong\cW_1(2)$
(Fig.~\ref{abb:linie}), which has projective dimension one. Its null-polarity
is the identity mapping. Two operators of $\vG\setminus\vK$ commute if, and
only if, their corresponding points are on a common line through the nucleus.

\begin{figure}[th]\unitlength0.4\textwidth
    \hfill
\begin{minipage}[t]{70mm}
\centering
    \includegraphics{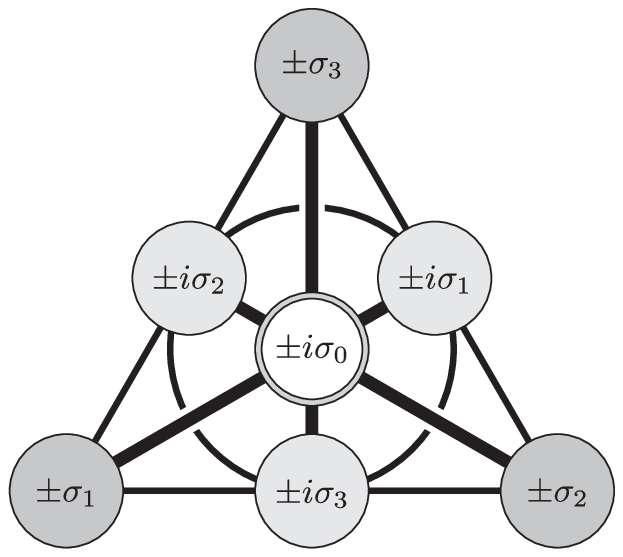}
  \caption{The f\/ine structure of the complex single-qubit Pauli group in terms of the
  Fano plane.}\label{abb:fano}
\end{minipage}\qquad
\begin{minipage}[t]{70mm}
    \centering
    \includegraphics{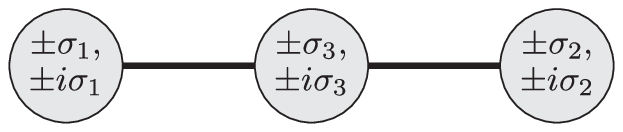}
    \caption{A coarser representation,~$\cW_1(2)$,  \emph{aka} the projection from the nucleus of
    the conic.}\label{abb:linie}
\end{minipage}
\hfill{}
\end{figure}

The normal subgroup $Z(\vG)$ satisf\/ies Conditions~\ref{c:el-abel} and
\ref{c:zentral}, but not \ref{c:a-quadrat}. The factor group $\vG/Z(\vG)$ has
$2^2$ elements; it gives rise to a two-dimensional symplectic vector space $V$
over $\GF(2)$ and the symplectic polar space $\cW_1(2)=\bP(V)$; see
Fig.~\ref{abb:linie}. The factor space $V_0/V_0^\perp$ from above and $V$ are
isomorphic (as symplectic vector spaces). Each point of $\cW_1(2)$ is totally
isotropic. We have no quadratic form on $V$. Two operators of $\vG\setminus
Z(\vG)$ commute if, and only if, their corresponding points are identical.
\end{example}

\begin{example}\label{exa:pauli64}
We exhibit the group comprising the Kronecker products
$i^\alpha\sigma_\beta\otimes\sigma_\gamma$ with $\beta,\gamma\in\{0,x,y,z\}$;
cf.\ (\ref{eq:pauli}). This group acts on the four-dimensional Hilbert space of
two qubits. In contrast to Example~\ref{exa:pauli16}, the symbol $\vG$ denotes
now this group of order $64$. In our terminology (with $p:=2$) we have
$Z(\vG)=\{\pm \sigma_0\otimes\sigma_0, \pm i\sigma_0\otimes\sigma_0\}$,
$\vG'=\vG^{(2)}=\vK=\{\pm\sigma_0\otimes\sigma_0\}$, and
$g=-\sigma_0\otimes\sigma_0$. Up to a change of dimensions, the situation here
completely parallels that of the preceding example:
\par
The factor group $\vG/\vK$ gives rise to a four-dimensional projective space
$\bP(V_0)$ over $\GF(2)$ and a non-degenerate quadric $\cQ_4(2)$. We are not
familiar with any neatly arranged picture of this projective space with its
$31$ points and $155$ lines. However, the $15$ points and $15$ singular lines
of $\cQ_4(2)$, together with its nucleus and several points/lines of its
ambient space, can be illustrated as in Fig.~\ref{abb:quadric}. There are
$15$ lines joining the nucleus $\bP(V_0^\perp)$ with the points of the quadric
$\cQ_4(2)$; these lines become the ``points'' of the factor space
$\bP(V/V^\perp)\cong\cW_3(2)$.

The factor group $\vG/Z(\vG)$ yields a four-dimensional symplectic vector space
$V$ and the symplectic polar space $\cW_3(2)$ with $\bP(V)$ as set of points.
It is depicted in Fig.~\ref{abb:doily} which is known as the
\emph{doily}\footnote{Another remarkable illustration of $\cW_3(2)$ exhibiting,
like the doily, a pentagonal cyclic symmetry is the so-called
\emph{Cremona--Richmond conf\/iguration}.}. We have no quadratic form on $V$. Two
operators of $\vG\setminus Z(\vG)$ commute if, and only if, their corresponding
points are on a totally isotropic line.

\begin{figure}[th]\unitlength0.4\textwidth
    \hfill
\begin{minipage}[t]{70mm}
\centering
    \includegraphics{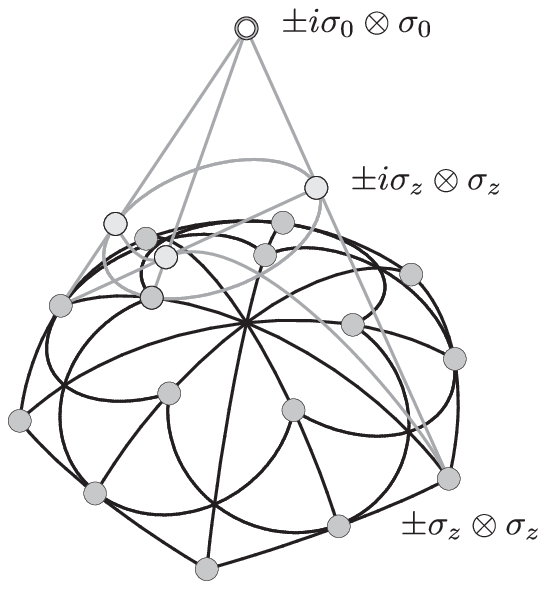}
  \caption{$\cQ_4(2)$, its nucleus, and a portion of its ambient space as the geometry behind the complex
  two-qubit Pauli group.}\label{abb:quadric}
\end{minipage}\qquad
\begin{minipage}[t]{70mm}
    \centering
    \includegraphics{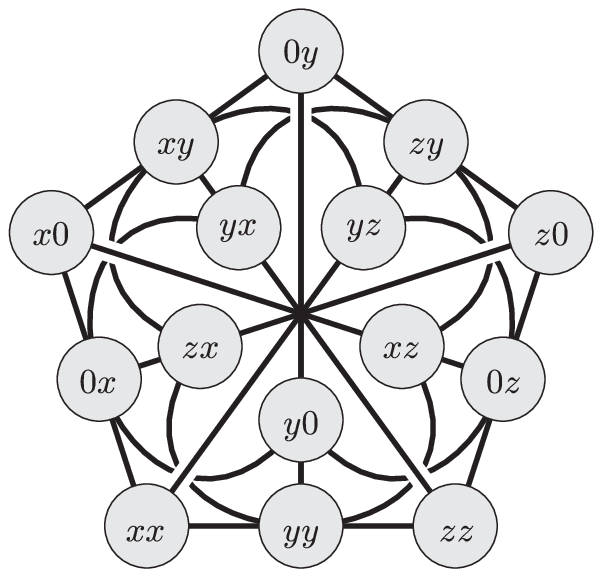}
    \caption{A coarser view in terms of~$\cW_3(2)$; $xy$ is a short-hand for
    $i^\alpha\sigma_x \otimes\sigma_y$, $\alpha \in \{0, 1$, $2, 3\}$, etc.}\label{abb:doily}
\end{minipage}
\hfill{}
\end{figure}

\end{example}

\begin{example}\label{exa:pauli8}
The real orthogonal matrices $\pm I$, $\pm X$, $\pm Y$, $\pm Z$, where
\begin{equation*}
    I:=
    \begin{pmatrix}
    1      & 0     \\
    0      & 1     \\
    \end{pmatrix},\qquad
    X:=
    \begin{pmatrix}
    0      & 1     \\
    1      & 0     \\
    \end{pmatrix},\qquad
    Y:=
    \begin{pmatrix}
    \phantom{-}0      & 1     \\
              -1      & 0     \\
    \end{pmatrix},\qquad
    Z:=
    \begin{pmatrix}
    1      & \phantom{-}0     \\
    0      &-1     \\
    \end{pmatrix},
\end{equation*}
constitute the \emph{real Pauli group} $\vG$. It acts on the Hilbert space
$\bR^2$ of a real single qubit. In our terminology (with $p:=2$) we have
$\vG'=\vG^{(2)}=\vK=Z(\vG)=\{\pm I\}$ and $g=-I$. Hence there is only one
possibility for factorisation, namely~$\vG/Z(\vG)$. This gives the symplectic
polar space~$\cW_1(2)$ based on the projective line over $\GF(2)$ which we
already encountered in Examp\-le~\ref{exa:pauli16}. However, now this space is
ref\/ined by an orthogonal polar space based on a hyperbolic quadric~$\cQ_1^+(2)$. The two points of this quadric represent those matrices in~$\vG\setminus Z(\vG)$ whose square is $I$ (i.e.\ symmetric matrices), the
remaining point corresponds to matrices in $\vG$ with square $-I$ (i.e.\
skew-symmetric matrices); see Fig.~\ref{abb:linie+q}.
\begin{figure}[th]\unitlength0.4\textwidth
    \hfill
\begin{minipage}[t]{70mm}
\centering
    \includegraphics{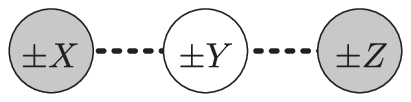}
  \caption{$\cW_1(2)$ and $\cQ_1^+(2)$ (shaded) of the real single-qubit Pauli group.}\label{abb:linie+q}
\end{minipage}\qquad
\begin{minipage}[t]{70mm}
\centering
    \includegraphics{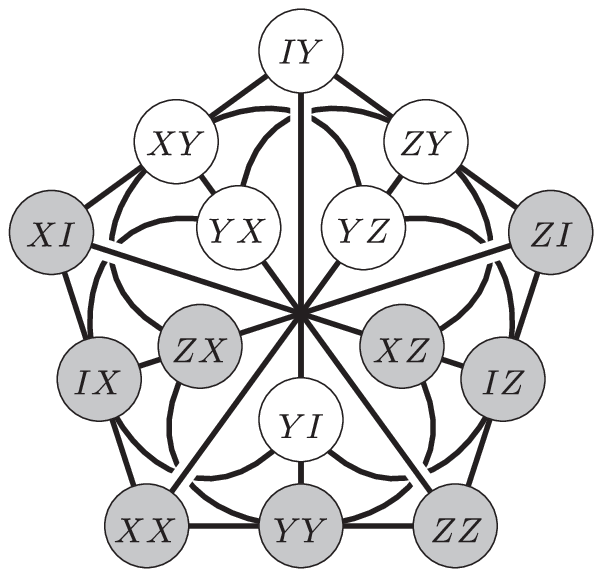}
    \caption{$\cW_3(2)$ and $\cQ_3^+(2)$ (shaded) of the real two-qubit Pauli group. $XY$ is a short-hand for
    $\pm X \otimes Y$, etc.}\label{abb:doily+q}
\end{minipage}
\hfill{}
\end{figure}
\end{example}

\begin{example}
Here we deal with the group comprising the Kronecker products of the matrices
from Example~\ref{exa:pauli8}. We change notation as now this group of order
$32$ is denoted by $\vG$. With $p:=2$ we have $\vG'=\vG^{(2)}=\vK=Z(\vG)=\{\pm
I\otimes I\}$. Up to a change of dimensions, the situation here completely
parallels that of the preceding example: The factor group $\vG/Z(\vG)$ gives
rise to the symplectic polar space $\cW_3(2)$ which is ref\/ined by an orthogonal
polar space based on a hyperbolic quadric $\cQ_3^+(2)$. The nine points of this
quadric represent matrices in $\vG\setminus Z(\vG)$ whose square is $I\otimes
I$ (i.e.\ symmetric matrices), the remaining points correspond to matrices in~$\vG$ which square to $-I\otimes I$ (i.e.\ skew-symmetric matrices); see
Fig.~\ref{abb:doily+q}.
\end{example}

\begin{example}
Finally, we mention the ($p=3$) case of \emph{two-qutrit} Pauli group (see also
\cite{planat+saniga-08a}). This group $\vG$ possesses $3^5$ elements, which can
be written in the form $\omega^a X^b Y^c \otimes X^d Y^e$, where $a, b, c, d, e
\in \{0, 1, 2\}$, $\omega$ is a primitive 3-rd root of unity, and $X$ and $Z$
are so-called shift and clock operators given by
\begin{equation*}
    \begin{pmatrix}
    0      & 0    & 1\\
    1      & 0    & 0\\
    0      & 1    & 0
    \end{pmatrix}
\qquad    \mbox{and}\qquad
    \begin{pmatrix}
    1      & 0     &0  \\
    0      &\omega &0  \\
    0      & 0     &\omega^2
      \end{pmatrix},
\end{equation*}
respectively (see, e.\,g., \cite{havlicek+saniga-07a,havlicek+saniga-08a}). Its factor group $\vG/Z(\vG)$ is of cardinality
$3^4 = 81$ and generates the symplectic polar space $\cW_{3}(3)$ of 40
points/lines, with 4 points on each line and, dually, 4~lines through each
point. This case is noteworthy in two crucial aspects. First, it is one of the
simplest instances where a single point of the associated polar space
represents not only a single operator (up to complex multiples), but
encompasses the \emph{two distinct powers\/} of an operator (up to complex
multiples). Second, it leads to far-reaching physical implications for the
so-called black-hole analogy (see, e.g., \cite{borsten+d+d+e+r-09z}). As per
the latter fact, it has recently been shown \cite{levay+s+v+p-08a} that the
$E_{6(6)}$ symmetric entropy formula describing black holes and black strings
in $D=5$ is intimately tied to the geometry of the generalised quadrangle
$\GQ(2,4)$, where $27$ black-hole charges correspond to the points and $45$
terms in the entropy formula to the lines of~$\GQ(2,4)$. And there exists a
very intimate connection between $\cW_{3}(3)$ and $\GQ(2,4)$
\cite{payne+thas-84a}. Given any point~$U$ of~$\cW_{3}(3)$, we can ``derive''
$\GQ(2,4)$ as follows. The points of~$\GQ(2,4)$ are all the points of~$\cW_{3}(3)$ not collinear with~$U$, whereas the lines of~$\GQ(2,4)$ are on the
one side the lines of~$\cW_{3}(3)$ not containing $U$ and on the other hand
hyperbolic lines through $U$ (natural incidence). Hence, this link between the
two f\/inite geometries not only unveils the mystery why $D=5$ black hole
solutions are related with qutrits, but knowing that each point of $\cW_{3}(3)$
comprises a couple ($p-1=3-1=2$) of elements of $\vG/Z(\vG)$, it also provides
a straightforward recipe for labelling the 45 members of the entropy formula in
terms of all elements of the two-qutrit Pauli group $\vG$.
\end{example}

Following these examples the interested reader should be able to f\/ind out the
symplectic and orthogonal polar spaces behind any (multiple-)qudit Pauli group
as long as the rank $d$ of the qudit is a prime number.

\subsection*{Acknowledgements}
This work was carried out in part within the ``Slovak-Austrian Science and
Technology Cooperation Agreement'' under grants SK 07-2009 (Austrian side) and
SK-AT-0001-08 (Slovak side), being also partially supported by the VEGA grant
agency projects Nos. 2/0092/09 and 2/7012/27. The f\/inal version was completed
within the framework of the Cooperation Group ``Finite Projective Ring
Geometries: An Intriguing Emerging Link Between Quantum Information Theory,
Black-Hole Physics, and Chemistry of Coupling'' at the Center for
Interdisciplinary Research (ZiF), University of Bielefeld, Germany. The authors
are grateful to Wolfgang Herfort (Vienna) for his suggestions.

\pdfbookmark[1]{References}{ref}
\LastPageEnding

\end{document}